\documentstyle[12pt,bezier]{article}
\begin{document}
\def \inbar{\vrule height1.5ex width.4pt depth0pt}
\def \xC{\relax\hbox{\kern.25em$\inbar\kern-.3em{\rm C}$}}
\def \xR{\relax{\rm I\kern-.18em R}}
\newcommand{\R}{\xR}
\newcommand{\C}{\xC}
\newcommand{\xZ}{Z \hspace{-.08in}Z}
\newcommand{\Z}{Z \hspace{-.08in}Z}
\newcommand{\xbe}{\begin{equation}}
\newcommand{\be}{\begin{equation}}
\newcommand{\xee}{\end{equation}}
\newcommand{\ee}{\end{equation}}
\newcommand{\xbea}{\begin{eqnarray}}
\newcommand{\bea}{\begin{eqnarray}}
\newcommand{\xeea}{\end{eqnarray}}
\newcommand{\eea}{\end{eqnarray}}
\newcommand{\xnn}{\nonumber}
\newcommand{\nn}{\nonumber}
\newcommand{\xkt}{\rangle}
\newcommand{\kt}{\rangle}
\newcommand{\xbr}{\langle}
\newcommand{\br}{\langle}
\newcommand{\xcun}{\mbox{\footnotesize${\cal N}$}}
\newcommand{\cun}{\mbox{\footnotesize${\cal N}$}}
\newcommand{\cum}{\mbox{\footnotesize${\cal M}$}}
\newcommand{\F}{{\cal F}}
\newcommand{\PP}{{\cal P}}
\newcommand{\Q}{{\cal Q}}

%
%

\title{On the Statistical Origin of Topological Symmetries}
\author{K.~Aghababaei Samani$ ^{a}$ and
A.~Mostafazadeh$ ^{b}$\\ \\
$ ^{a}$~Institute for Advanced Studies in Basic Sciences,\\
45195-159 Gava Zang, Zanjan, IRAN\thanks{E-mail address:
samani@iasbs.ac.ir}\\
$ ^{b}$~Department of Mathematics, Ko\c{c} University,\\
Rumelifeneri Yolu, 80910 Sariyer, Istanbul, TURKEY\thanks{E-mail address:
amostafazadeh@ku.edu.tr}}
\date{ }
\maketitle

\begin{abstract}
We investigate a quantum system possessing a parasupersymmetry of order $2$, an orthosupersymmetry of order 
$p$, a fractional supersymmetry of order $p+1$, and topological symmetries of type $(1,p)$ and $(1,1,\cdots,1)$. We 
obtain the corresponding symmetry generators, explore their relationship, and show that they may be expressed  in 
terms of the creation and annihilation operators for an ordinary boson and orthofermions of order $p$. We give a realization 
of parafermions of order~$2$ using orthofermions of arbitrary order~$p$, discuss a $p=2$ parasupersymmetry 
between $p=2$ parafermions and parabosons of arbitrary order, and show that every orthosupersymmetric system 
possesses topological symmetries. We also reveal a correspondence between the orthosupersymmetry of order $p$
and the fractional supersymmetry of order $p+1$.
\end{abstract}


\baselineskip=24pt

\section{Introduction}
During the past two decades supersymmetry has been widely used in a number of problems in quantum mechanics
\cite{susy,review,junker}. This has been one of the main reasons for the development and the study of various 
generalizations of supersymmetric quantum mechanics (SQM). Among the best known examples of these 
generalizations are the parasupersymmetric quantum mechanics (PSQM) \cite{ru-sp,be-de,du-ma-sp,psusy-p,khare}, the 
fractional supersymmetric quantum Mechanics (FSQM) \cite{fsusy,durand,az-ma}, and the orthosupersymmetric quantum 
mechanics (OSQM) \cite{ortho}. Recently, we have introduced a class of generalizations of SQM that unlike the above
mentioned generalizations

 share the topological properties of supersymmetry in the sense that these 
symmetries involve a set of integer-valued topological invariants. We have termed these symmetries {\em Topological 
Symmetries}. In Refs.~\cite{p1,p2}, we develop the concept of a topological symmetry (TS), obtain the underlying 
operator algebra, and show that the algebras of SQM, PSQM of order~2, and FSQM of arbitrary order are special cases 
of the algebras of different types of TSs. An important difference between our approach to TSs and the standard 
approach to SQM and PSQM is that we do not define TS as a generalization of the symmetries of an oscillator 
consisting of two degrees of freedom with specific statistics. The purpose of the present article is to seek for such a 
statistical interpretation of TS.

It is well-known that algebras of SQM \cite{dewitt,susy}, PSQM \cite{ru-sp,psusy-p,khare}, and OSQM \cite{ortho} 
have been originally obtained as symmetries of an oscillator consisting of a bosonic degree of freedom and a fermionic, 
a parafermionic, or an orthofermionic degree of freedom, respectively. Other realizations of SQM in a pure 
parafermionic system, a pure parabosonic system, and a parafermi-parabose oscillator have also been discussed in the 
literature \cite{ply1,ply2,ijmpa96c}. Moreover, a $k$-fermionic realization of PSQM is offered in 
Ref.~\cite{daoud-kibler-99}.

The algebra of FSQM has been originally obtained as a simple but rather ad hoc generalization of the algebra of SQM 
that was mainly influenced by the ideas of $q$-deformed bosons \cite{fsusy,durand,az-ma}. A somewhat concrete justification 
for the basic algebraic relations used in FSQM is given in Ref.~\cite{flipov}. The idea of finding a realization of FSQM 
using an oscillator consisting of a boson and another particle has been pursued in 
Refs.~\cite{az-ma,daoud-kibler-01}.  

In this paper we consider a quantum mechanical system whose Hamiltonian has a parasupersymmetry of order~2,
a fractional supersymmetry of order $p+1$, an orthosupersymmetry of order $p$, and TSs of type $(1,p)$ and 
$(1,1,\cdots,1)$. We show that the generators of these symmetries have a statistical interpretation in terms of bosons 
and orthofermions. 

The organization of the paper is as follows. In Section 2, we review topological symmetries. In Section 3, we survey the 
algebras of PSQM, OSQM, and FSQM. In Section 4, we use a concrete system to investigate the statistical origin of 
these symmetries and their relationship. In Section 5, we show that every OSQM system possesses TSs of type $(1,p)$ 
and $(1,1,\cdots,1)$, and construct the generators of TSs in terms of the generators of OSQM. In Section 6, we give a 
summary of our results and present our concluding remarks.

\section{Topological Symmetries}

A topological symmetry is a generalization of supersymmetry in the sense that it allows for the introduction of certain 
topological invariants. These are the analogues of the Witten index of SQM \cite{witten-82}. Here we give the
definition and the operator algebra of the topological symmetries. For more details we refer the reader to 
Ref.~\cite{p2}.

A quantum system is said to possess a $\Z_n$-graded (uniform) topological symmetry (UTS) of type 
$(m_1,m_2,\cdots,m_n)$ iff the following conditions are satisfied. 
	\begin{itemize}
	\item[1.] The quantum system is $\Z_n$-graded. This means that the Hilbert space ${\cal H}$ of the quantum 
	system is the direct sum of $n$ of its (nontrivial) subspaces ${\cal H}_\ell$, and its Hamiltonian  has a complete 
	set of eigenvectors with definite {\it color} or grading. (A state is said to have a definite color $c_\ell$ iff it  
	belongs to ${\cal H}_\ell$);
	\item[2.] The  energy  spectrum is nonnegative;
	\item[3.] Every positive energy eigenvalue $E$ is $m$-fold degenerate, and the corresponding eigenspaces are 
	spanned by $m_1$ vectors of color $c_1$, $m_2$ vectors of color $c_2$, $\cdots$, and $m_n$ vectors of color 
	$c_n$.
	\end{itemize}
For a system with these properties we can introduce a set of integer-valued topological invariant, namely
    \be
    \Delta_{ij}:= m_in_j^{(0)} -  m_jn_i^{(0)},
    \label{1.3}
    \ee
where $i,j=1,\cdots,n$ and $n_\ell^{(0)}$ denotes the number of zero-energy states of color $c_\ell$, \cite{p2}. 
Note that the TS of type $(1,1)$ coincides with supersymmetry and $\Delta_{11}$ yields the Witten index.

As shown in Ref.~\cite{p2}, one can obtain the underlying operator algebra of TSs. For a ($\Z_2$-graded)
UTS of type $(1,p)$ one finds the algebra
    \bea
    &&[H,\Q]=0\;,
    \label{ts1}\\
    &&\{\Q^2,\Q^\dagger\}+\Q\Q^\dagger\Q=2H\Q\;,
    \label{ts2}\\
    &&\Q^3=0\;,
    \label{ts3}\\
   &&[H,\tau]=\{\tau,\Q\}=0\;,
   \label{ts4}\\
   && \tau^2=1\;,~~ \tau^\dagger=\tau\;,
    \label{t2}
    \eea
where $H$ is the Hamiltonian of the system, $\Q$ is the generator of UTS, and $\tau$ is the grading operator. We 
can also express these algebraic relation in terms of Hermitian generators:
    \be
    Q_1:=\frac{1}{\sqrt{2}}\:(\Q+\Q^\dagger)\;~~~
    {\rm and}~~~Q_2:=\frac{-i}{\sqrt{2}}\:(\Q-\Q^\dagger)\;.
    \label{2.1}
    \ee
This yields
   \be
   Q_1^3=Q_1H\; ~~~{\rm and}~~~Q_2^3=Q_2H\;.
   \label{2.2}
   \ee
The algebra of ($\Z_n$-graded) UTS of type $(\underbrace{1,1,\cdots,1}_{n~{\rm times}})$ is given by
    \bea
   &&\Q^n=K\;, \label{2.3}\\
    &&Q_1^n +M_{n-2} Q_1^{n-2} +\cdots
    =({1\over \sqrt 2})^n (2K)\;,
    \label{2.4}\\
    &&Q_2^n +M_{n-2} Q_2^{n-2} +\cdots
    =({1\over \sqrt 2})^n (i^n+ (-i)^n)K \;,
    \label{2.5}\\
   &&[\tau,\Q]_q=0\;.
   \label{2.6}
    \eea
where $M_i$s and $K$ are Hermitian operators commuting with all other operators, and $\tau$ is the grading operator 
satisfying
    \bea
    \tau^n&=&1\;,
    \label{tn1}\\
    \tau^\dagger&=&\tau^{-1}\;,
    \label{tn2}\\
    \left[H,\tau\right]&=&0\;.
    \label{tn3}
    \eea
Here $[.,.]_q$ stands for {\it $q$-commutation} defined by $[O_1,O_2]_q:=O_1O_2-qO_2O_1$, and 
$q:=e^{2\pi i/n}$. One should note that this is not the most general form of the algebra of UTS of type 
$(1,1,\cdots,1)$, but this form suffices for the purposes of this article.

\section{Para, Ortho and Fractional Supersymmetry}

In this section we survey the algebras of quantum systems with parasupersymmetry of order $2$, orthosupersymmetry 
of order $p$, and fractional supersymmetry of order~$n$.

\subsection{Algebra of the PSQM of order $2$}
A $p=2$ PSQM system is defined by the following algebra \cite{ru-sp}.
    \bea
    &&[H,\PP]=0\;,
    \label{ps1}\\
    &&\{\PP^2,\PP^\dagger\}+\PP\PP^\dagger\PP=4H\PP\;,
    \label{ps2}\\
    &&\PP^3=0\;.
    \label{ps3}
   \eea
Here $\PP$ is the generator of the parasupersymmetry. Comparing Eqs.~(\ref{ps1}) -- (\ref{ps3}) with 
Eqs.~(\ref{ts1}) -- (\ref{ts3}), we see that the algebra of $p=2$ PSQM and UTS of type $(1,p)$ are identical.
This does not, however, 
imply that any $p=2$ parasupersymmetric system possesses a UTS of type~$(1,p)$. The key observation is that the 
defining properties of UTS of type~$(1,p)$ lead to the algebra~(\ref{ps1}) -- (\ref{ps3}), but this algebra does not 
imply the defining degeneracy structure of the UTS of type~$(1,p)$.\footnote{The set of $p=2$ parasupersymmetric 
systems which do satisfy the defining conditions of UTS of type~$(1,p)$ form a proper subset of all the $p=2$ 
parasupersymmetric systems, \cite{ijmps97}.}

\subsection{Algebra of the OSQM of order $p$}
The algebra of an OSQM of order $p$ is given by 
   \bea
   &&[H,\Q_\alpha]=0\;,
   \label{os1}\\
   &&\Q_\alpha \Q_\beta^\dagger +\delta_{\alpha \beta} \sum^p_{\gamma=1}
   \Q_\gamma^\dagger \Q_\gamma=2\delta_{\alpha \beta}H\;,
   \label{os2}\\
   &&\Q_\alpha \Q_\beta=0\;,
   \label{os3}
   \eea
where $\Q_\alpha$, with $\alpha=1,2,\cdots,p$, are the generators of the orthosupersymmetry \cite{ortho}.

\subsection{Algebra of the FSQM of order $n$}
FSQM of order $n$ is defined by the relation
   \be
   \F^n=H\;,
   \label{fs1}
   \ee
where $\F$ is the generator of fractional supersymmetry \cite{fsusy,durand}. Note that $\F$ need not be Hermitian. 
Furthermore, as suggested by the algebra (\ref{2.3}) -- (\ref{2.6}) of the $\Z_n$-graded UTS of type $(1,1,\cdots,1)$,
any system having a $\Z_n$-graded UTS of type $(1,1,\cdots,1)$ also has a fractional supersymmetry of order $n$ 
\cite{p2}. 

\section{Statistical Origin of TS in a Simple Toy Model}

Consider a quantum system with the Hamiltonian
   \be
   H=a^\dagger a+\sum^p_{\gamma=1}c_\gamma^\dagger c_\gamma\;,
   \label{4.1}
   \ee
where $a$ is the annihilation operator for a bosonic degree of freedom satisfying
	\be
	[a,a^\dagger ]=1\;,
	\label{4.2}
	\ee
and $c_\gamma\;$, with $\gamma=1,\cdots,p$, are annihilation operators of orthofermions of order $p$ satisfying 
\cite{ortho}
   \bea
   &&c_\alpha c_\beta^\dagger +\delta_{\alpha \beta} \sum^p_{\gamma=1}
   c_\gamma^\dagger c_\gamma=\delta_{\alpha \beta}\;,
   \label{of1}\\
   &&c_\alpha c_\beta=0\;.
   \label{of2}
   \eea
A simple representation of $c_\alpha$ is given by the $(p+1)\times (p+1)$ matrices
   \be
   [c_\alpha]_{ij}=\delta_{i,1}\delta_{j,\alpha+1}\;,~~ i,j=1,\cdots,p+1\;.
   \label{4.3}
   \ee
In this representation, the Hamiltonian of the system reads
   \be
   H={\rm diag}(a^\dagger a,\underbrace{aa^\dagger ,\cdots,a a^\dagger }_{p~{\rm times}})\;,
   \label{4.4}
   \ee
where `diag$(\cdots)$' stands for `diagonal matrix with diagonal elements $\cdots$'.

It is easy to check that the Hamiltonian~(\ref{4.1}) and the operator
   \be
   \Q_\alpha={\sqrt 2}a^\dagger c_\alpha\;,~~\alpha=1,\cdots,p
   \label{4.5}
   \ee
satisfy the algebra (\ref{os1}) -- (\ref{os3}) of OSQM.\footnote{Note that we postulate relative 
bosonic statistics between bosons and orthofermions, i.e., $[a,c_\alpha]=0$ for all $\alpha$.}
   
Next, consider the operator
   \be
   \PP=\frac{2}{\sqrt p}\left(a \sum^r_{j=1}c^\dagger_{\gamma_j} +a^\dagger \sum^p_{j=r+1}
   c_{\gamma_j}\right)\;,
   \label{4.6}
   \ee
where $(\gamma_1,\cdots\gamma_p)$ is an arbitrary permutation of $(1,\cdots,p)$ and $r$ is an integer between 
$1$ and $p-1$. It turns out that the Hamiltonian (\ref{4.1}) and the operator $\PP$ fulfil the algebra 
(\ref{ps1}) -- (\ref{ps3}) of $p=2$ PSQM. Clearly, $\PP$ is determined by the choice of the permutation 
$(\gamma_1,\cdots\gamma_p)$. Therefore this system has many parasupersymmetries of order 2.

In the next section, we shall prove that in general every orthosupersymmetric quantum mechanical system possesses a
$p=2$ parasupersymmetry. Parasupersymmetry was originally introduced as a symmetry of a 
bosonic-parafermionic oscillator \cite{ru-sp}. However, there are also other statistical interpretations for PSQM. An 
example is a realization of the parasupersymmetry algebra using $k$-fermions \cite{daoud-kibler-99}.
Another example is a parabose-parafermi parasupersymmetry of the Hamiltonian:
   \be
   H=\frac{1}{2}\{ \tilde a^\dagger ,\tilde a\}+\frac{1}{2}[b^\dagger ,b]
   \label{4.61}
   \ee
where $\tilde a$ is the annihilation operator of a paraboson of arbitrary order $p=1,2,\cdots$ satisfying 
   \be
   [\tilde a^2, \tilde a^\dagger ]=2\tilde a\;,
   \label{pba}
   \ee
and $b$ is the annihilation operator of a parafermion of order $2$ satisfying
   \bea
   &&b^3=0\;,
   \label{pfa1}\\
   && b^\dagger b^2 +b^2 b^\dagger=bb^\dagger b=2b\;.
   \label{pfa2}
   \eea
The Hamiltonian~(\ref{4.61}) and the operator
   \be
   \PP=\tilde a b^\dagger\;,
   \label{psch}
   \ee
fulfil the algebra of PSQM of order $2$, namely Eqs.~(\ref{ps1}) -- (\ref{ps3}).\footnote{Here we postulate relative
bosonic statistics between parafermions and parabosons, i.e., $[\bar a,b]=[\bar a,b^\dagger]=0$.}

It is well-known that one can construct parafermionic and parabosonic operators using the Green's ansatz \cite{green}. 
The above relationship between $p=2$ parasupersymmetry and orthosupersymmetry suggests the following 
construction of $p=2$ parafermions in terms of orthofermions:
   \be
   b= \sum^r_{j=1}\alpha_j c^\dagger_{\gamma_j} + \sum^p_{j=r+1}
   \alpha_j c_{\gamma_j}\;,
   \label{5.2}
   \ee
where $1\leq r<p$ and $\alpha_i$ satisfy
   \be
   \sum^r_{j=1}|\alpha_j|^2= \sum^p_{j=r+1}
   |\alpha_j|^2=2\;.
   \label{5.3}
   \ee
For example, we may set
    \be
    \alpha_i=\left\{
    \begin{array}{ccc}
    {\sqrt {2 \over r}}&{\rm for} & 1 \le i \le r\\
    {\sqrt {2 \over {p-r}}}&{\rm for}& r+1 \le i \le p.\\
    \end{array}\right.
    \label{5.4}
    \ee
It is not difficult to show that $b$ as given by Eq.~(\ref{5.2}) satisfies the algebra of $p=2$ parafermions, namely 
Eqs.~(\ref{pfa1}) and (\ref{pfa2}).

Furthermore, we can also check that the Hamiltonian (\ref{4.1}) with the operator $\Q=\PP/{\sqrt 2}$ satisfy both the 
algebra (\ref{ts1}) -- (\ref{ts4}) and the defining conditions for the UTS of type $(1,p)$. The corresponding grading 
operator is given by
   \be
   \tau=(-1)^N,~~~{\rm where}~~~N:=\sum^p_{\alpha=1}N_\alpha\;,
   \label{4.7}
   \ee
and $N_\alpha$ are orthofermionic number operators:
    \be
    N_\alpha:=c^\dagger_\alpha c_\alpha\;.
    \label{4.71}
    \ee

Next, we note that the Hamiltonian~(\ref{4.61}) has a fractional supersymmetry of order $p+1$ generated by 
   \be
   \F=ac_1^\dagger +c_2^\dagger c_1+\cdots+c_p^\dagger c_{p-1}+a^\dagger c_p\;.
   \label{4.8}
   \ee
Finally, this Hamiltonian has, in addition, a $\Z_{p+1}$-graded UTS of type $(1,1,\cdots,1)$ with the same generator 
as $\F$ and the grading operator
   \be
   \tau=q^{\cal N},~~~{\rm where}~~~{\cal N}:=\sum^p_{\alpha=1}\alpha N_\alpha\;,
   \label{4.9}
   \ee
the operator $K$ of Eq.~(\ref{2.3}) is identified with the Hamiltonian $H$, the operators $M_i$ of 
Eqs.~(\ref{2.4}) and (\ref{2.5}) are given in terms of $H$ according to
        \be
        M_{n-2k}=(-1)^k\left[{1 \over {2^k}} {{n-k-1} \choose k}+
        {1 \over {2^{k-1}}}{{n-k-1} \choose {k-1}}H\right]\;,
        \label{4.10}
        \ee
and $\left(\begin{array}{c}a\\b\end{array}\right):=\frac{a!}{b!(a-b)!}$.

\section{Relation between General OSQM and TSs}

In this section we show that orthosupersymmetric systems always possess TSs of type $(1,p)$ and $(1,1,\cdots,1)$.

Consider an arbitrary orthosupersymmetric Hamiltonian $H$ satisfying Eqs.~(\ref{os1}) -- (\ref{os3}) and let
 \be
   \Q:=\frac{1}{\sqrt p}\,\left(\sum^r_{j=1}\Q^\dagger_{\gamma_j} + \sum^p_{j=r+1}
   \Q_{\gamma_j}\right)\;,
   \label{5.1}
   \ee
where $(\gamma_1,\cdots,\gamma_p)$ is a permutation of $(1,\cdots,p)$. Then, it can be easily checked that 
$\Q$ is a generator for a TS of type $(1,p)$.\footnote{This is actually to be expected for the system has 
the required degeneracy structure \cite{ortho}.} The $\Z_2$-grading of the Hilbert space is established 
via the grading operator 
   \be
   \tau=(-1)^{({\Q_1 \Q_1^\dagger \over 2H})}=1-\frac{\Q_1 \Q_1^\dagger}{H}.
   \label{grts2}
   \ee
Furthermore, the Hamiltonian (\ref{os2}) can be expressed in terms of the generators $\Q_\gamma$ according to
   \be
   H=\frac{1}{2}(\Q_1 \Q_1^\dagger + \sum^p_{\gamma=1}
   \Q_\gamma^\dagger \Q_\gamma)\;,
   \label{hamex}
   \ee
where we have used the fact that $\Q_1 \Q_1^\dagger=\cdots=\Q_p \Q_p^\dagger$.

Next, we use the generators of the orthosupersymmetry $\Q_\gamma$ to construct a generator for the TS of type 
$(1,1,\cdots,1)$ for this system, namely
   \be
   \Q=\Q_1^\dagger +\Q_2^\dagger \Q_1 +\cdots + \Q_p^\dagger\Q_{p-1} +\Q_p.
   \label{xts1}
   \ee
This operator together with the Hamiltonian (\ref{hamex}) and the grading operator
   \be
   \tau=(q)^{\sum^p_{\alpha=1}\frac{\alpha \Q_\alpha^\dagger \Q_\alpha}{2H}}=
   1+\frac{1}{2H}\sum^p_{\alpha=1}\Q_\alpha^\dagger \Q_\alpha(q^\alpha-1)
   \label{grtsn}
   \ee
satisfy the algebra (\ref{2.3}) -- (\ref{tn3}) of the $\Z_{p+1}$-graded TS of type $(1,1,\cdots,1)$. The 
operators $K$ and $M_{n-2k}$ appearing in this algebra are given by
   \bea
   &&K=(2H)^p\;,
   \label{k} \\
   &&M_{n-2k}=(-2)^k H^{2k-1}\left[ {{n-k-1} \choose k}H+   
	{{n-k-1} \choose {k-1}}\right]\;.       
   \label{mnk}
   \eea
In particular, $K$ may be viewed as the Hamiltonian of a fractional supersymmetric system. Therefore, to 
every orthosupersymmetric system of order $p$, there corresponds a fractional supersymmetric system of order $p+1$.

\section{Concluding Remarks}
In this article, we have explored the statistical origins of certain topological symmetries. We have obtained an 
interpretation of topological symmetries of type $(1,p)$ and $(1,1,\cdots,1)$ in terms of certain symmetries between 
bosons and orthofermions. We have given a realization of $p=2$ parafermions in terms of orthofermions of arbitrary 
order, and revealed a novel relationship between fractional supersymmetry and orthosupersymmetry. We have also 
shown that every orthosupersymmetric system has topological symmetries of type $(1,p)$ and $(1,1,\cdots,1)$. Our 
results have potential application in the construction of quantum systems possessing topological symmetries and 
having a topologically nontrivial configuration space. The study of these systems is of particular importance for a better
understanding of the mathematical meaning of the topological invariants $\Delta_{ij}$.

\end{document}